      [1999/12/01 v1.4c Il Nuovo Cimento]
\documentclass{cimento}

\renewcommand{\IN}[4]{\textit{#1}~\textbf{#2}~(#3)~#4}

\usepackage{graphicx}  
\title{Associated production of top quarks and charged Higgs bosons at next-to-leading order}
\author{C.~Weydert\from{ins:x}}
\instlist{\inst{ins:x} LPSC, UJF-CNRS/IN2P3-INPG, 53 avenue des Martyrs 38026 Grenoble, France}
\PACSes{\PACSit{12.38.Bx,12.60.Fr}{}}
\begin{document}

\maketitle

\begin{abstract}
Accurate predictions for both signal and background events at the LHC are of paramount importance in order to confirm even the smallest deviations from Standard Model predicitions. Next-to-leading order Monte Carlo event generators are an essential tool to reach that goal. Concerning the charged Higgs boson, NLO calculations of the production cross section already exist. Reiterating the calculation using a subtraction formalism enables us to implement the cross section into Monte Carlo generators, which can then be used by experiments.
\end{abstract}

\section{Introduction}

MC@NLO~\cite{ref:mcnlo} is widely used by the experiments at the LHC for top production cross section estimations and will thus also be the default generator for charged Higgs boson production, implying that this implementation had to be absolute priority~\cite{ref:caro}.  However, MC@NLO is known to have some drawbacks for experimental use. So far, MC@NLO is interfaced only with HERWIG, and using another parton shower (PS) is not straightforward. In many cases the preferred PS is Pythia, since it is easily tunable to data. Another complication, which arises when using MC@NLO, is the possibility to encounter negatively weighted events due to the particular combination of Monte Carlo generator and next-to-leading order (NLO) calculation. This can be an issue if the experimental analysis is performed via trained multivariate techniques, which are unable to handle events with negative weights. A remedy to both issues can be found in the use of POWHEG~\cite{ref:pwhg}.\\
Providing an implementation in two independent generators is very useful. One must never forget that, since event generators are complicated, man-coded computer codes, they are never free of bugs. At least two estimations from different codes can give a clearer hint as to potential errors and should be performed whenever possible. 


\section{Charged Higgs boson production at next-to-leading order (NLO) at the LHC}
The first NLO calculations of charged Higgs boson production in association with a top quark have been performed almost a decade ago~\cite{ref:zhu, ref:til}. We have used them to check our results. A NLO cross section is very important to evaluate production cross sections, but also for tests of the universality of the charged Higgs boson couplings~\cite{ref:Dean}.\\
The following paragraph sketches the difficulties encountered when adding up different contributions of the NLO cross section.
Initial-state collinear divergencies are cancelled with counterterms coming from the NLO PDFs. The remaining divergencies cancel between the virtual $ \sigma^V$ and the real cross section $\sigma^R$, so that the total NLO cross section is finite. Although the sum is finite, we need to separate these pieces in order to perform the integration, since they involve different final state phase spaces,
\begin{equation}\label{e.without}
	 \sigma^{NLO} = \int_{3} d\sigma^R + \int_{2} d \sigma^V.
\end{equation}
Various solutions to this problem have been proposed. The most popular one in the early days of NLO calculations was the aforementioned phase-space slicing method.  A somewhat different approach is the so-called Catani-Seymour dipole formalism~\cite{ref:cs}. The general philosophy of this method is that in order to have a numerically integrable cross section, an auxiliary term $d \sigma^A$ is defined. This terms has two special features, namely it exhibits the same pole structure as $d \sigma^R$ and can thus act as a local counterterm, and it is analytically integrable over the singular one-particle subspace. The right hand side of Eq.~(\ref{e.without}) can thus be rewritten as
\begin{equation} \label{e.with}
	 \sigma^{NLO} = \int_{3} \left[ \left( d\sigma^R \right)_{\epsilon=0}  -  \left( d\sigma^A \right)_{\epsilon=0}  \right]
		+ \int_{2} \left[  d\sigma^V  + \int_1 d \sigma^A  \right]_{\epsilon=0},
\end{equation}
and the integrations can now be performed over finite quantities. \\
More information on this and on the implementation into MC@NLO can be found in~\cite{ref:caro}. 

\section{Implementation in POWHEG}
This section briefly summarizes ongoing work concerning the implemenation of the NLO cross section calculation into POWHEG. The following normalised plots show a comparison between the pure NLO calculation and the POWHEG result when coupled to HERWIG for a charged Higgs boson mass of $m_{H^{-}}= 300$ GeV, a coupling to fermions (up-type $u_i$ and down-type $d_j$) of type II $\left[\mathcal{L}\propto H^{+} \bar{u}_i (\frac{m_{u_i}}{\tan \beta} P_L + m_{d_j} \tan \beta P_R ) d_j\right]$  with $P_{R/L}=(1\pm \gamma^5)/2$ and $\tan \beta = 30$, and a center-of-mass energy of the LHC of $\sqrt{S}= 14$ TeV.
\begin{figure} 
\centering
\includegraphics[scale=0.22]{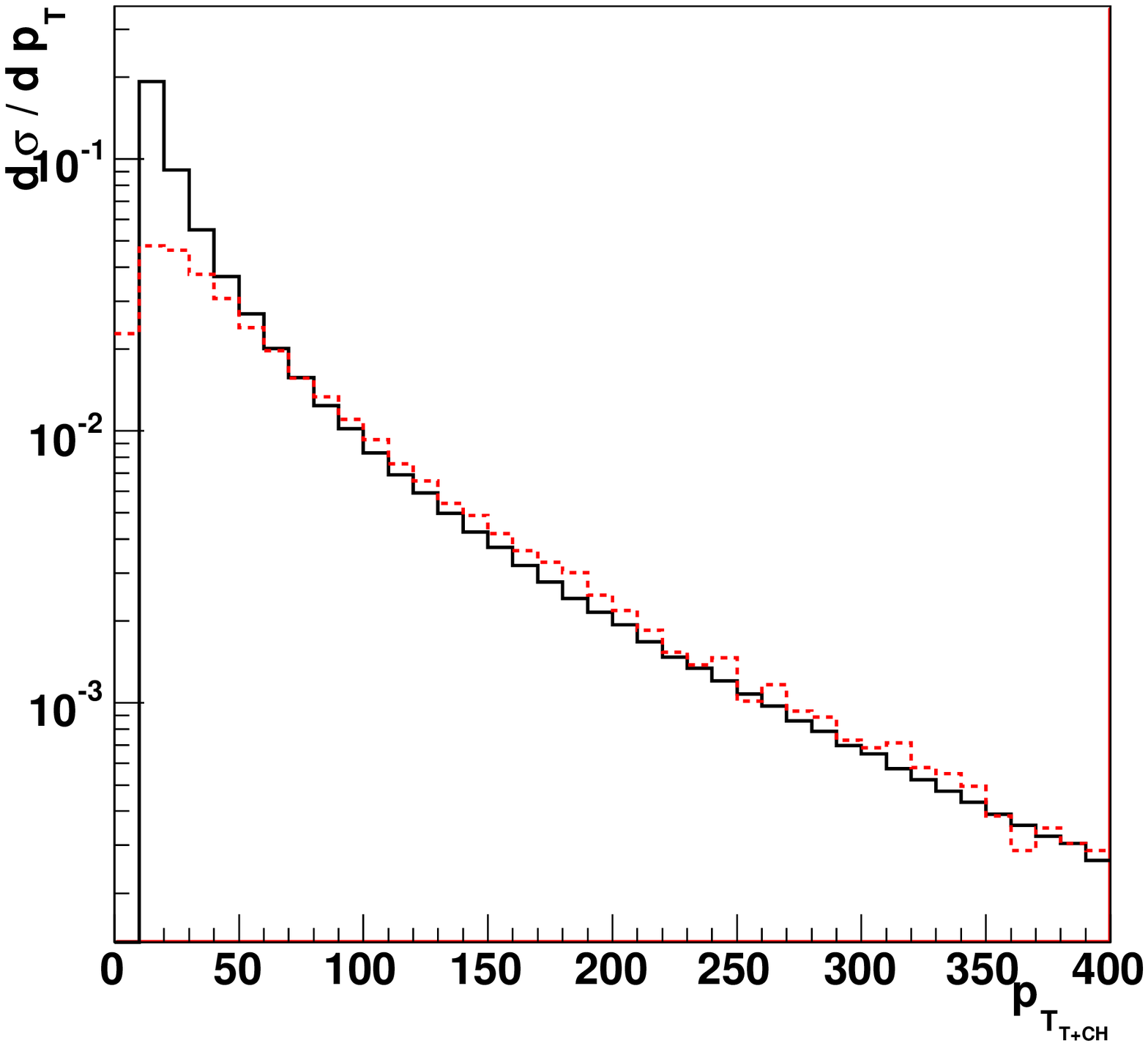}
\includegraphics[scale=0.22]{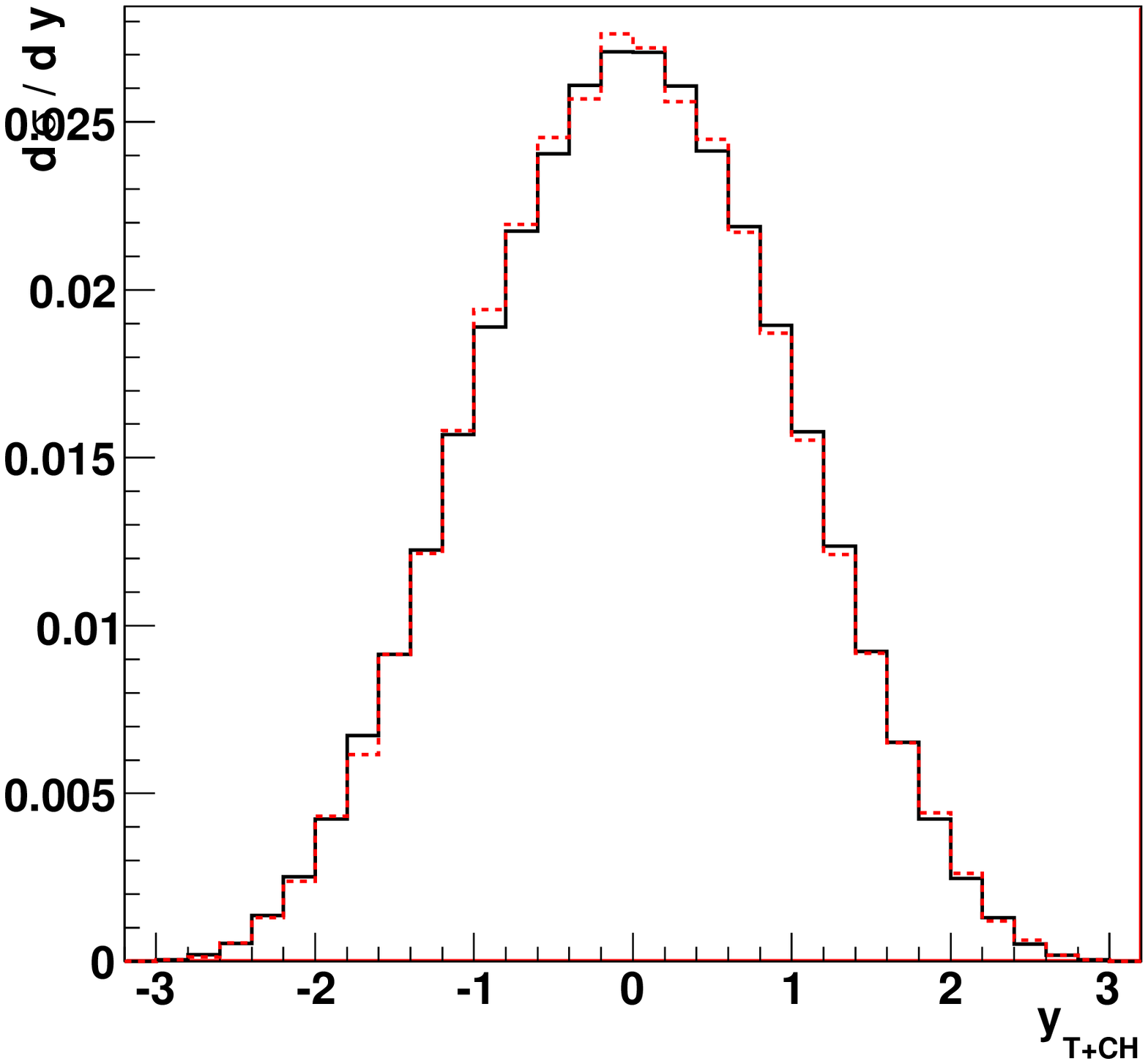}
\includegraphics[scale=0.22]{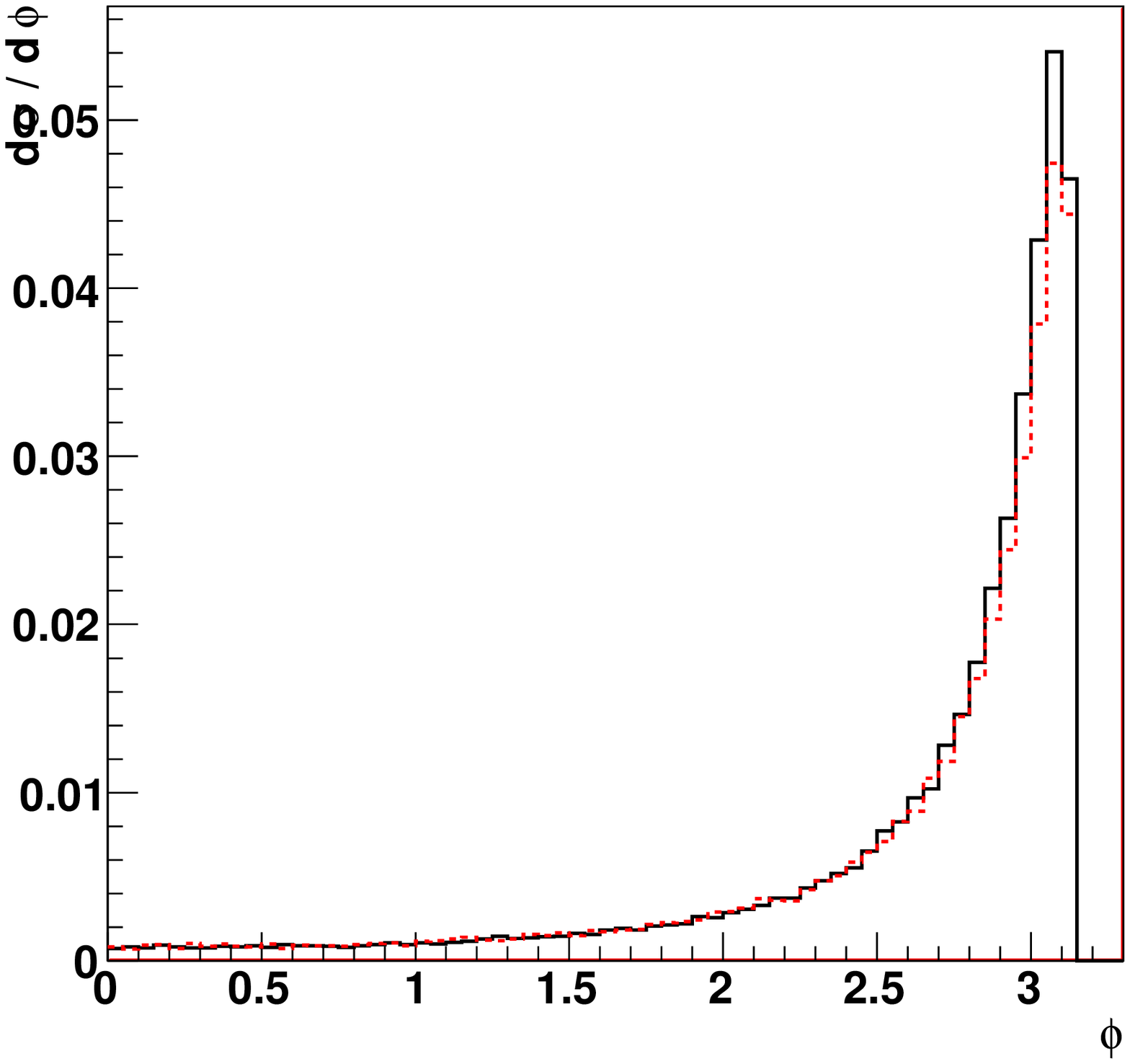}
\caption{\label{fig:all} Comparison between the purely NLO calculation (solid line) and POWHEG+HERWIG (dashed line) for the pair transverse momentum $\vert \vec{p}_{T_{top}}+\vec{p}_{T_{H^{-}}} \vert, $ the pair rapidity $ y_{top+H^{-}}$ and the azimuthal angle $\phi$ between the top quark and the charged Higgs boson.}
\end{figure}
The left plot of figure~\ref{fig:all} displays the transverse momentum distribution of the top quark and the charged Higgs boson. The pure NLO curve is negative for the first bin and then reaches very high values. This typical behaviour is seen to be smoothened by the PS in POWHEG. A resummed calculation would also be similar to the PS behavior. The second plot shows the rapidity distribution of the top quark and the charged Higgs boson pair. Agreement between the pure NLO calculation and the POWHEG output coupled to the PS can be seen. Finally, the last plot displays the azimuthal angle between the top quark and the charged Higgs boson. Again, the PS regularizes the behavior of the NLO calculation at $\phi = \pi.$

\section{Outlook}
The implementation in POWHEG for large charged Higgs boson masses is complete and the low-mass $(m_{CH}<m_{t})$ case is in progress. However, there are still some issues that could be explored further. \\
Our calculation relies on specific simplifications, as for example, neglecting the bottom quark mass with respect to other kinematic variables. Keeping the bottom quark massless is referred to as the five-flavor scheme. A compraison with the four-flavor scheme calculation~\cite{ref:kraemer} would be of great interest. An analogous comparison has already been performed for single top production in the $t$-channel~\cite{ref:camp}. Also, a proper resummation could be done for this process in order to compare the low-$p_T$ regions with the Monte Carlo output.

\end{document}